\documentclass[12pt]{article}

\usepackage[pdftex,usenames,dvipsnames]{color}
\usepackage{graphics}
\usepackage{a4wide}
\usepackage{amsfonts}

\newcommand{\nit}{\noindent}

\newcommand{\np}{\newpage}

\newcommand{\vs}[1]{\vspace{#1 ex}}
\newcommand{\hs}[1]{\hspace{#1 em}}
\newcommand{\bflr}{\begin{flushright}}
\newcommand{\eflr}{\end{flushright}}
\newcommand{\bc}{\begin{center}}
\newcommand{\ec}{\end{center}}
\newcommand{\ben}{\begin{enumerate}}
\newcommand{\een}{\end{enumerate}}

\newcommand{\be}{\begin{equation}}
\newcommand{\ee}{\end{equation}}
\newcommand{\ba}{\begin{array}}
\newcommand{\ea}{\end{array}}

\newcommand{\bg}{\beta}
\newcommand{\gam}{\gamma}

\begin{document}

\bflr
NIKHEF/2014-050
\eflr

\bc
{\large {\bf Neutrinos}}\\
\vs{2}

{\large {\bf a window on new physics}}
\vs{3}

J.W. van Holten$^a$ 
\vs{3}

Nikhef, Amsterdam NL 
\vs{1}

and Lorentz Insitute 
\vs{1}

Leiden University, Leiden NL
\ec
\vs{5}

\nit
{\footnotesize 
{\bf Abstract} \\
This paper reviews some aspects of the physics of neutrinos, in particular neutrino masses and the issue of
Dirac versus Majorana neutrinos. The see-saw mechanism is described and it is argued that the Majorana
nature of neutrinos can be tested by measuring the invisible decays of the Higgs particle, as its decay into
neutrinos is determined by their Yukawa couplings, i.e.\ the Dirac masses, rather than the physical Majorana 
masses. The measurement would allow us to probe the scale $M$ of the large Majorana masses for right-handed
singlet neutrinos. The optimal machine for performing such a measurement would be a future electron-positron 
collider. \\
\\
This paper is prepared for the Netherlands Journal of Physics (Nederlands Tijdschrift voor Natuurkunde) and 
is written in Dutch. }

\vfill
\footnoterule
{\footnotesize $^a$ e-mail: v.holten@nikhef.nl}

\np
\nit
{\bf In de afgelopen vijf jaar is met de protonenbundels van de LHC zeer succesvol onderzoek
uitgevoerd. Het Higgs-deeltje werd gedetecteerd, de massa en spin van het deeltje zijn gemeten en 
een groot aantal verschillende vervalwijzen van het deeltje zijn in kaart gebracht. Meting van de 
koppeling aan neutrino's zou ons belangrijke informatie over nu nog onbekende fysica kunnen 
verschaffen.}
\vs{1}

\nit
Het verval van Higgs-deeltjes in fotonen, weergegeven in fig.\ 1, is een van de manieren om deze 
deeltjes te identificeren en hun massa te bepalen. Het bestaan van Higgs-deeltjes verschaft antwoord
op belangrijke vragen uit de subatomaire fysica. In de eerste plaats is het mechanisme gevonden 
achter de breking van de symmetrie van de elektro-zwakke wisselwerkingen van quarks en leptonen. 
Die symmetrie zou ervoor moeten zorgen dat de $W$- en $Z$-deeltjes die bij deze wisselwerkingen 
worden uitgewisseld net zo massaloos zijn als het foton. Maar het alomtegen\-woordige 
Brout-Englert-Higgs-veld (BEH) geeft deze deeltjes een rustenergie, dus massa. Het Higgs-deeltje 
is een manifestatie van het bestaan van dit veld. 

Het BEH-veld draagt ook bij aan de massa van fermionen, zoals quarks en geladen leptonen: 
het elektron, het muon en het $\tau$-deeltje. De laatste drie krijgen zelfs al hun massa via dit 
mechanisme, en voor de zware quarks (zoals de $b$- en $t$-quarks) is dat in belangrijke mate 
het geval. De sterkte van de interactie met het BEH-veld is ongelijk voor verschillende deeltjes
en wordt beschreven door een karakteristieke parameter, de Yukawa-koppeling. Bijgevolg zijn 
de massa's van geladen leptonen of zware quarks ook verschillend: ze zijn evenredig met hun 
Yukawa-koppeling. Maar de Yukawa-koppelingen bepalen ook hoe makkelijk het Higgs-deeltje 
vervalt in een deeltje en antideeltje van de betreffende soort. Een Higgs-deeltje zal daardoor vaak in 
massieve $b$- en $\bar{b}$-quarks vervallen, maar slechts zelden in een licht $e^+e^-$-paar. Dit is 
precies wat de experimenten bij de LHC sinds de ontdekking van het Higgs-deeltje hebben laten zien. 
\vs{2}

\nit
{\bf Na het Standaardmodel} \\ 
Het Standaardmodel van de subatomaire fysica, dat de kleurkrachten tussen quarks en de 
elektromagnetische en zwakke wisselwerkingen van quarks en leptonen {\nobreak beschrijft,} is 
daarmee grotendeels compleet. De vraag dringt zich dan op of er op nog kleinere afstand\-schalen 
nieuwe verschijnselen optreden. Dat is aannemelijk omdat de zwaartekracht, die in het 
Standaardmodel geen rol speelt, in de kosmologie en de astrofysica een centrale plaats inneemt. 
Volgens de meest voor de hand liggende scenario's wordt de quantumzwaartekracht relevant op 
afstanden van $10^{-33}$ cm of kleiner, de Planck-schaal. Om zulke afstanden te verkennen met 
versnellers zou je deeltjes met een energie van $10^{19}$ GeV moeten kunnen maken. Tussen 
de zwakke wissel\-werkingen in het Standaardmodel, die spelen op een schaal van $10^{-15}$ cm, 
en de Planck-schaal ligt echter een groot gebied dat nog niet in kaart gebracht is. 
\vs{2}

\nit
{\bf Neutrino's} \\
De meest begaanbare weg naar dit onbekende gebied zou wel eens kunnen lopen via de allerlichtste 
deeltjes uit de fermionfamilies van het Standaardmodel: de neutrino's. Neutrino's zijn in het 
Standaardmodel in verscheidene opzichten een speciaal geval. Ze zijn minstens een miljoen keer 
lichter dan elektronen, en dragen geen elektrische lading of kleurlading. Dientengevolge zijn hun 
interacties met de rest van de materie zo gering dat ze door vrijwel niets tegen te houden zijn, en 
ons zelfs vanuit het midden van de zon met de snelheid van het licht bereiken. 

Net als andere fermionen in het Standaardmodel komen neutrino's voor in drie soorten, die door de 
zwakke wisselwerkingen elk met een van de drie geladen leptonen verbonden zijn: $\nu_e$, 
$\nu_{\mu}$ en $\nu_{\tau}$. Het is zeker dat neutrino's een massa bezitten, want verscheidene 
experimenten hebben neutrino-oscillaties waargenomen: processen waarbij neutrino's van de ene 
soort spontaan in een andere soort overgaan. Dat is mogelijk omdat de massieve neutrino's een 
tijdafhankelijke quantummechanische superpositie van de drie wissel\-werkingstypen zijn. Zo'n  
tijdsafhankelijkheid ontstaat alleen als de massa's verschillend zijn. Uit de waarnemingen volgt 
een menging en een massaspectrum als in fig.\ 2. Omdat ze zo extreem licht zijn is de absolute 
waarde van de neutrinomassa's echter onbekend; er is alleen een bovenlimiet van ongeveer 
1 eV/$c^2$. Die limiet doet vermoeden dat neutrinomassa's misschien een andere oorsprong 
hebben dan de Yukawa-koppeling aan het BEH-veld. 

Dat vermoeden wordt gevoed door een kwestie aangaande de aard van neutrino's.  Als elektrisch en 
kleurneutrale deeltjes zouden neutrino's namelijk wel eens hun eigen anti\-deeltjes kunnen zijn: 
Majorana-fermionen. Dit in tegenstelling tot de andere fermionen in het standaardmodel, die in paren 
van tegengesteld geladen deeltjes en antideeltjes voorkomen: Dirac-fermionen.
\vs{5}

{\fbox{\parbox{5.7in}{
\bc {\em Leptongetal} 
\ec
\nit
Een belangrijke regel in de quantumelectrodynamica, een onderdeel van het standaardmodel, is dat 
uit fotonen alleen paren van een negatief geladen lepton en een positief geladen antilepton kunnen 
ontstaan. Ook kunnen geladen leptonen alleen paarsgewijs annihileren tot fotonen. 
Het netto aantal geladen leptonen (aantal deeltjes minus aantal antideeltjes) is dan voor en na de
reactie hetzelfde. De zwakke wisselwerkingen kunnen roet in het eten gooien, omdat een
$W^-$-deeltje uiteen kan vallen in een antineutrino en een geladen lepton, en een $W^+$-deeltje in 
een neutrino en een geladen antilepton. Als neutrino's en antineutrino's verschillend zijn, dan is nog
steeds het aantal geladen leptonen en neutrino's minus het aantal geladen antileptonen en 
antineutrino's behouden. Dit staat bekend als het behoud van leptongetal, waarbij (anti)neutrino's als
ongeladen (anti)leptonen tellen. Als er echter geen onderscheid is tussen neutrino's en antineutrino's, 
dan gaat de bereke\-ning niet meer op en wordt het behoud van leptongetal geschonden in zwakke 
wisselwerkingen. \\
}}}

\np
\nit
{\bf Neutrinomassa's}\\
Een probleem met Majorana-neutrino's is, dat hun massa niet door Yukawa-koppeling aan het 
BEH-veld kan worden veroorzaakt; een Yukawa-koppeling aan het BEH-veld is alleen mogelijk 
voor Dirac-fermionen. Toch moet ook de massa van Majorana-neutrino's een dynamische oorsprong 
hebben: het is een fundamentele eigenschap van deeltjes met zwakke wisselwerkingen dat ze intrinsiek 
massaloos zijn. Dat is te be\-grij\-pen aan de hand van $\bg$-verval. De standaardvorm van $\bg$-verval is 
de overgang van een neutron in een proton via emissie van een virtueel $W^-$-deeltje; dit produceert dan 
een elektron en een anti-neutrino.  Ook kan een proton soms overgaan in een neutron via emissie van een 
virtueel $W^+$-deeltje, dat in een positron en een neutrino uiteenvalt. Het bijzondere van de zwakke 
interactie via $W$-deeltjes is, dat elektronen of neutrino's er altijd met een linkshandige polarisatie 
uitkomen, terwijl positronen of antineutrino's een rechts\-handige polarisatie meekrijgen. Dit betekent dat 
$W$-deeltjes uiteenvallen in deeltjes met spin antiparallel aan de bewe\-gingsrichting en antideeltjes met 
spin parallel aan de bewe\-gingsrichting. 

Nu ligt de gepolariseerde produktie van massieve deeltjes in $\bg$-verval niet voor de hand, omdat 
de polarisatierichting geen fundamentele eigenschap is. Immers, voor een waarnemer die met het 
deeltje meebeweegt staat het deeltje stil, en is het concept van een spinrichting parallel of antiparallel 
aan de bewegingsrichting zinloos; en ten opzichte van een waar\-nemer die sneller is dan het deeltje 
keert de bewegingsrichting van het deeltje om, terwijl de spin niet omklapt. Massaloze deeltjes vormen 
echter een uitzondering: die bewegen altijd met de lichtsnelheid en het is niet mogelijk ze in te halen of 
met ze mee te bewegen; voor zulke deeltjes is de pola\-risatie {\em wel} een fundamentele eigenschap. 
In het Standaard\-model zijn zowel de $W$- en $Z$-deeltjes als de fermionen die daar zwakke interacties 
mee hebben daarom van zichzelf massaloos. Hun massa ontstaat als een dynamisch effect. Voor 
Dirac-deeltjes is dit de koppeling aan het BEH-veld.

Als neutrino's ook Dirac-deeltjes zijn, en hun massa afkomstig is van de Yukawa-koppeling aan het
BEH-veld, dan moet die Yukawa-koppeling uiterst klein zijn. We kunnen dit met een paar getallen illustreren. 
Voor de grootste neutrinomassa nemen we een re\-presentatieve waarde van 0.15 eV/$c^2$. In verhouding
tot de massa van het elektron: $m_e = 0.5$ MeV/$c^2$, is dit neutrino nog tien keer lichter dan het 
elektron in vergelijking met het $t$-quark ($m_t = 170$ GeV/$c^2$):
\be
\frac{m_{\nu}}{m_e} = \frac{m_e}{10 m_t} \simeq 3 \times 10^{-7}.
\label{a}
\ee 
Dit zou impliceren dat de Yukawa-koppeling van neutrino's meer dan zes ordes van grootte kleiner is
dan die van het elektron, en dertien ordes van grootte kleiner dan die van het $t$-quark. Zo'n kleine 
Yukawa-koppeling van neutrino's betekent ook dat het verval van een Higgs-deeltje in een 
$\nu\bar{\nu}$-paar, hoewel theoretisch mogelijk, praktisch nooit voorkomt.

Neutrino's kunnen alleen Dirac-deeltjes zijn als er ook rechtshandige neutrino's en linkshandige 
antineutrino's bestaan. Dat moeten ware spookdeeltjes zijn zonder zwakke wissel\-werkingen. 
Ze kunnen niet worden gemaakt in $\bg$-verval. Hun enige mogelijk\-heid tot interactie is 
de uiterst zwakke Yukawa-koppeling aan het BEH-veld, waaruit de massa voortkomt. 
\vs{2}

\nit
{\bf Majorana-neutrino's}\\
Als een neutrino echter een Majorana-deeltje is kan dit heel anders liggen. Een scenario voor 
Majorana-neutrino's werd ruim 30 jaar geleden voorgesteld door Peter Minkowski [3] in Zwitserland en 
door Murray Gell-Mann, Pierre Ramond en Richard Slansky [4] in de VS. Ook in hun voorstel zijn er 
rechtshandige spookneutrino's die geen zwakke wisselwerkingen met $W$- en $Z$-deeltjes hebben, 
maar wel een Yukawa-koppeling aan het BEH-veld. De vorming van Dirac-neutrino's wordt in dit scenario 
echter voorkomen door nog onbekende dynamica bij zeer kleine afstanden waardoor de rechtshandige 
neutrino's een grote massa krijgen die hen tot zware Majorana-deeltjes maakt. De wisselwerking met de 
linkshandige standaardneutrino's via het BEH-veld zorgt dan voor een menging tussen de gewone en de 
zware neutrino's, wat uiteindelijk een stel lichte en een stel zware Majorana-neutrino's oplevert. De zware 
mengneutrino's houden daarbij hun grote massa $M$, maar de lichte mengneutrino's krijgen een massa 
$m_{\nu}$ die onder de massa $m_D$ ligt die ze als Dirac-deeltjes via de Yukawa-koppeling zouden krijgen: 
\be
\frac{m_{\nu}}{m_D} = \frac{m_D}{M}.
\label{b}
\ee
Hoe groter de massa $M$ uit de nieuwe wisselwerkingen bij de onbekende afstandschaal, hoe kleiner 
de massa $m_{\nu}$ van de waargenomen neutrino's. Daarom heet dit het {\em see-saw} (wip-wap)
mechanisme. Door de uitdrukkingen (\ref{a}) en (\ref{b}) te vergelijken zien we dat er realistische 
neutrinomassa's uitkomen als $m_D$ ongeveer gelijk aan de elektronmassa is en $M$ het tienvoudige 
van de top-massa. Maar als de massa $M$ bij voorbeeld tienmiljoen keer zo groot is: $M = 10^8\; m_t$, 
dan is $m_D$ vergelijkbaar met de massa van de $b$-quarks. Omdat de massa $m_D$ aangeeft hoe 
sterk de neutrino's aan het BEH-veld en aan het Higgs-deeltje koppelen, zou dit betekenen dat het 
Higgs-deeltje net zo vaak in neutrino's als in $b$-quarks zou moeten vervallen. 
\vs{2}

\nit
{\bf Kunnen we Majorana-neutrino's detecteren?}\\
De conclusie uit dit scenario is, dat de vervalkans van een Higgs-deeltje in neutrino's niet samenhangt 
met de fysische massa-schaal $m_{\nu}$, maar met de bijdrage van de Dirac-massa $m_D$ daaraan. 
Een bepaling van $m_D$ levert ons dan bij bekende $m_{\nu}$ direct  de massa $M$ van de zware 
neutrino's, en daarmee de schaal waarbij de nieuwe wisselwerkingen optreden. Het is daarvoor niet
nodig energie\"{e}n van de orde $Mc^2$ met een versneller te bereiken. Wel is het nodig te kunnen 
meten hoe vaak het Higgs-deeltje in neutrino's vervalt. Het meest ideale experiment zou zijn elektron-positron 
botsingen bij de energie van de Higgs-massa (126 GeV) te bestuderen, waarbij Higgs-deeltjes in rust 
gemaakt worden en hun verval in allerlei kanalen relatief makkelijk te meten is. Maar het is ook niet
uitgesloten dat de experimenten met de LHC informatie kunnen opleveren waaruit we iets over de 
Yukawa-koppelingen van  neutrino's aan het BEH-veld kunnen leren. Als die koppelingen meetbaar 
zijn, is het Majorana-karakter van de lichte neutrino's onontkoombaar en zou er voor het eerst een 
venster op onbekende fysica ver voorbij het Standaardmodel geopend worden. 
\vs{2}

\np
\nit
Literatuur 
\vs{1}

\nit
\begin{tabular}{ll}
$[1]$ & The ATLAS collaboration, {\em Measurement of the Higgs boson mass from the $H \rightarrow \gam \gam$} \\
 & {\em and $H \rightarrow Z Z_*  \rightarrow 4l$ channels with the ATLAS detector using 25 fb$^{-1}$ of pp} \\
 & {\em  collision data};  Phys.\ Rev.\ D90 052004 (2014) \\
$[2]$ & R.\ de Adelhart Toorop. {\em A flavour of family symmetries in a family of flavour models} \\
 & Proefschrift, Univ.\ Leiden (2012) \\
$[3]$ & P.\ Minkowski, {\em $\mu \rightarrow e \gam$ at a rate of one out of $10^9$ muon decays?} \\
    & Phys.\ Lett.\ B67 (1977), 421 \\ 
$[4]$ & M.\ Gell-Mann, P.\ Ramond en R.\ Slansky, {\em Complex spinors and unified theories} \\
 & in: {\em Supergravity}, eds.\ P.\ van Nieuwenhuizen en D.Z.\ Freedman \\
 &  (North-Holland, 1979), 315 \\
\end{tabular}

\np

\bc
\scalebox{0.55}{\includegraphics{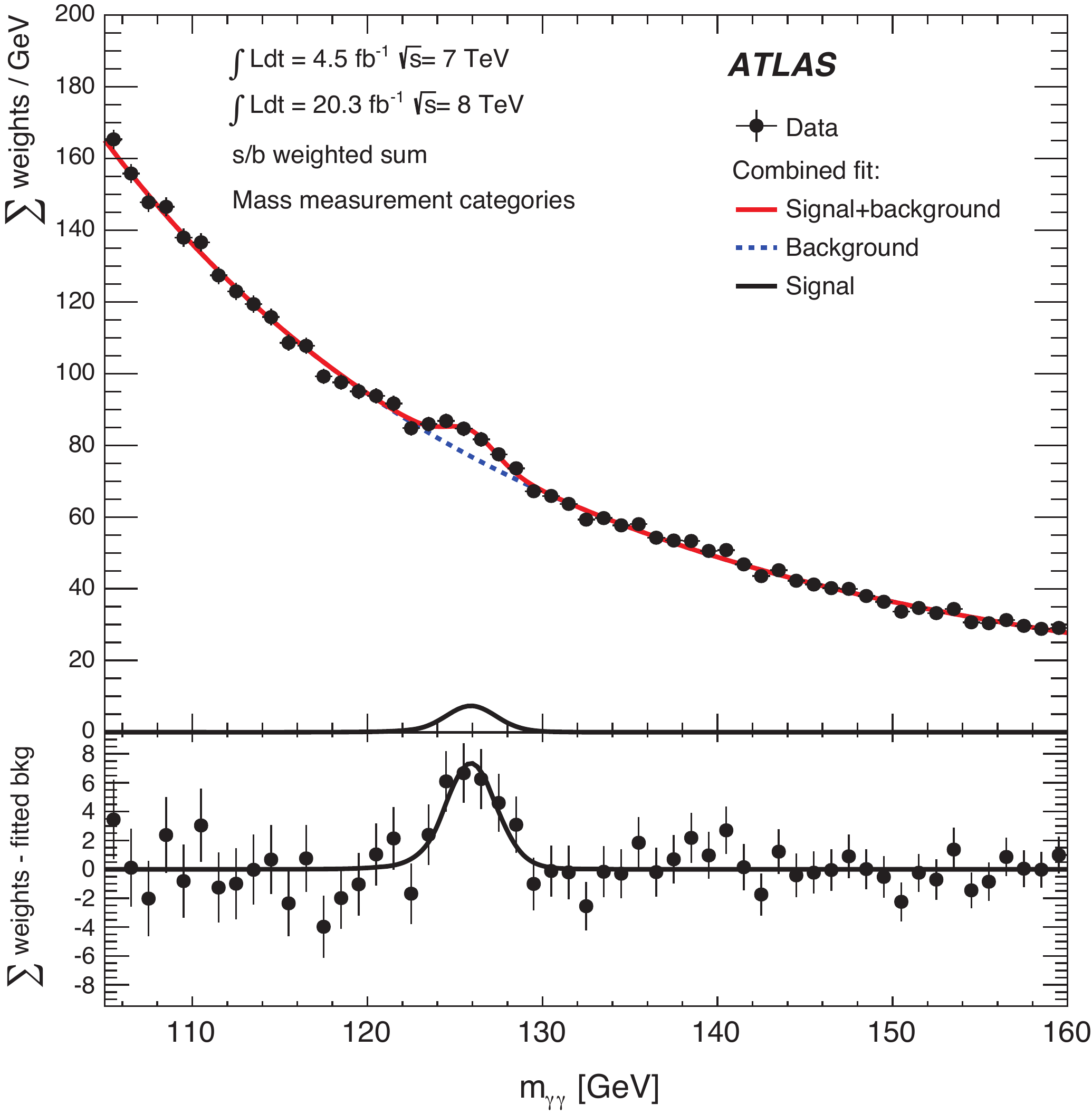}}

{\footnotesize{
\begin{tabular}{ll} Fig.\ 1: & Productie van 2 hoog-energetische fotonen in proton botsingen gemeten door \\
  & het ATLAS experiment $[1]$; de piek bij 126 GeV is de bijdrage van het verval \\ 
  & van Higgs-deeltjes. 
\end{tabular} }}
\vs{5}

\scalebox{0.25}{$\hs{5}$ \includegraphics{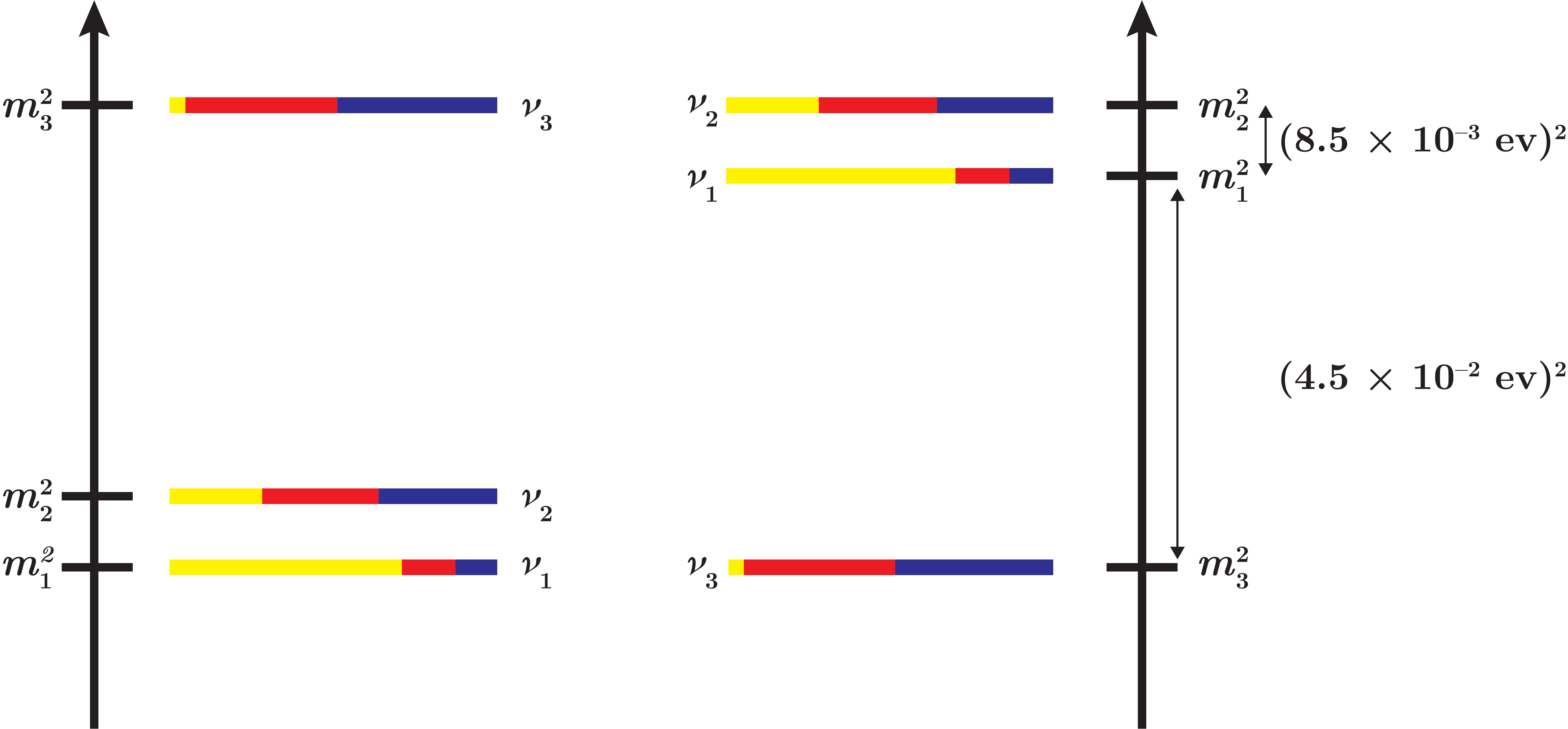}}

{\footnotesize{
\begin{tabular}{ll}
Fig.\ 2: & Massa spectrum van neutrino's; de massaverschillen zijn gemeten, de absolute waarde \\
 & is onbekend. Er zijn twee mogelijkheden aangegeven om de massa's te ordenen. De \\ 
 & kleuren representeren het aandeel van de verschillende interactietypen $(\nu_e, \nu_{\mu}, \nu_{\tau})$ aan \\ 
 & de neutrino's met massa $(m_1,m_2,m_3)$. Voor een overzicht, zie b.v.\ ref.\ $[2]$. 
\end{tabular} }}

\ec

\end{document}